\documentclass[%
 reprint,
 amsmath,amssymb,
 aps,
]{revtex4-1}

\usepackage{graphicx}
\usepackage{float}
\usepackage[export]{adjustbox} 
\usepackage{dcolumn}
\usepackage{bm}
\usepackage{braket}
\usepackage{amsthm}
\usepackage{caption}
\usepackage{subcaption}
\usepackage{amsmath}
\usepackage{amsthm}
\usepackage{amssymb}
\usepackage{bm}
\usepackage{braket}
\usepackage{amsfonts}
\usepackage{array}
\usepackage[thinlines]{easytable}
\usepackage{color}

\mathchardef\mhyphen="2D

\makeindex
\begin{document}

\title{Magnetic dipole excitations based on the relativistic nuclear energy density functional}%

\author{G. Kru{\v z}i{\' c}}
 \email{goran.kruzic@ericsson.com}
 \affiliation{Department of Physics, Faculty of Science, University of Zagreb, Bijeni{\v c}ka c. 32, HR-10000, Zagreb, Croatia}
 \affiliation{Research department, Ericsson - Nikola Tesla, Krapinska 45, HR - 10000, Zagreb, Croatia}

\author{T. Oishi}
 \email{toishi@phy.hr}
 \affiliation{Department of Physics, Faculty of Science, University of Zagreb, Bijeni{\v c}ka c. 32, HR-10000, Zagreb, Croatia}

\author{D. Vale}
\email{denivalesq@gmail.com}
 \affiliation{Department of Physics, Faculty of Science, University of Zagreb, Bijeni{\v c}ka c. 32, HR-10000, Zagreb, Croatia}

\author{N. Paar}%
 \email{npaar@phy.hr}
 \affiliation{Department of Physics, Faculty of Science, University of Zagreb, Bijeni{\v c}ka c. 32, HR-10000, Zagreb, Croatia}
%


\newcommand{\blu}[1]{\textcolor[rgb]{0.0,0.0,1.0}{\bf #1}}
\newcommand{\rosso}[1]{\textcolor[rgb]{1.0,0.0,0.0}{\bf #1}}

\newcommand{\bi}[1]{\ensuremath{\boldsymbol{#1}}}
\newcommand{\unit}[1]{\ensuremath{\mathrm{#1}}}
\newcommand{\oprt}[1]{\ensuremath{\hat{\mathcal{#1}}}}
\newcommand{\abs}[1]{\ensuremath{\left| #1 \right|}}

\newcommand{\crc}[1] {c^{\dagger}_{#1}}
\newcommand{\anc}[1] {c_{#1}}
\newcommand{\crb}[1] {a^{\dagger}_{#1}}
\newcommand{\anb}[1] {a_{#1}}

\newcommand{\slashed}[1] {\not\!{#1}} 

\def \beq{\begin{equation}}
\def \eeq{\end{equation}}
\def \beqa{\begin{eqnarray}}
\def \eeqa{\end{eqnarray}}

\def \bir{\bi{r}}
\def \ubir{\bar{\bi{r}}}
\def \bip{\bi{p}}
\def \ubip{\bar{\bi{r}}}

\begin{abstract}
{\noindent
Magnetic dipole (M1) excitations build not only a fundamental mode of nucleonic transitions, but they are also relevant for nuclear astrophysics applications. We have established a theory framework for description of M1 transitions 
based on the relativistic nuclear energy density functional. For this purpose the relativistic quasiparticle
random phase approximation (RQRPA) is established using density dependent point coupling interaction DD-PC1,
supplemented with the isovector-pseudovector interaction channel in order to study unnatural parity transitions.
The introduced framework has been validated using the M1 sum rule for core-plus-two-nucleon systems, and
employed in studies of the spin, orbital, isoscalar and isovector M1 transition strengths, that relate to the electromagnetic probe, in magic 
nuclei $^{48}$Ca and $^{208}$Pb, and open shell 
nuclei $^{42}$Ca and $^{50}$Ti. In these systems, the isovector spin-flip M1 transition is dominant, mainly 
between one or two spin-orbit partner states.  It is shown that pairing 
correlations have a significant impact on the centroid energy and major peak position of the M1 mode.  
The M1 excitations could provide an additional constraint to improve nuclear energy density functionals in the future studies.
}
\end{abstract}

\maketitle


\section{\label{sec:INTRO} Introduction}
Electromagnetic excitations in finite nuclei represent one of the most important probes of relevance in nuclear structure and dynamics, as well as in nuclear astrophysics. 
In particular, various aspects of magnetic dipole (M1) mode have been considered both in experimental and theoretical studies \cite{Richter02, 1995Richter, Richter01, Kneissl01, Pietralla01,Shizuma2008,2015Goriely, 2016Goriely} due to its relevance for diverse nuclear properties associated e.g., to unnatural-parity states, spin-orbit splittings and tensor force effects.
Specifically, M1 spin-flip excitations are analog of Gamow-Teller (GT) transitions, meaning that, at the operator level, 
the dominant M1 isovector component is the synonym
to the zeroth component of GT transitions, and can serve as probe for calculations of inelastic neutrino-nucleus cross section \cite{Langanke01, Langanke02}. 
This process is hard to measure but it is essential in supernova physics, as well as in the r-process nucleosynthesis calculations~\cite{Langanke2008,Loens2012,2015Goriely,2016Goriely}. 
The isovector spin-flip M1 response is also relevant for applications related to the design of
nuclear reactors~\cite{Chadwick2011}, for the understanding of single-particle properties, spin-orbit interaction, and shell closures from stable nuclei 
toward limits of stability~\cite{Otsuka2005,Otsuka2010,Nesterenko01,Nesterenko02,Tselyaev01}, as well
as for the resolving the problem of quenching of the spin-isopin response 
in nuclei that is necessary for reliable description of double beta decay 
matrix elements~\cite{Vergados2012}. 
In deformed nuclei, 
another type of M1 excitations has been extensively studied, 
known as scissors mode, where the orbital part of M1 operator plays a dominant role in a way that protons and neutrons oscillate with opposite phase around the core \cite{Arteaga01, Arteaga02, Richter01, Richter03, Otsuka01, Bohle01, Bohle02, Schwengner01, Balbutsev01, Repko01}. 

In any nuclei undergoing experimental investigation, 
there are simultaneously present $E\lambda$ and $M\lambda$ multipole excitations, 
where the electric dipole (E1) and electric quadrupole (E2) responses \cite{2019OP, Dang01, Yoshida01, Terasaki01, Ebata01, Stetcu01} dominate over M1 response \cite{Tamii01, Fujita, 2016Birkhan, Laszewski, Laszewski02, Laszewski03, Laszewski04, Raman01, Raman02}. 
Thus, it is a rather challenging task to measure M1-related observables in a whole energy range. 
Even for the nuclides accessible by experiments, their full information 
on the M1 response has not been complete. 

The M1 transitions have been studied in various theoretical approaches. 
Various aspects of the M1 mode have been investigated in the shell model \cite{Richter03, Otsuka2010, Loens2012, 2018Sieja, Otsuka2005, Langanke01, 1969Maripuu}, including, e.g.,
scissors and unique-parity modes \cite{Richter03},
tensor-force effect \cite{Otsuka2005}, 
low-energy enhancement of radiation \cite{2018Sieja}, 
and the analogy with neutrino-nucleus scattering \cite{Langanke01, Loens2012}.
The M1 energy-weighted sum rule has been discussed from a perspective of the spin-orbit energy \cite{1963Kurath}. The Landau-Migdal interaction has been one of the relevant topics in studies of M1 excitations \cite{1991Migli, 1993Kam}.
In order to reproduce a large fragmentation of the experimental M1 strength, the importance of including complex couplings going beyond the RPA level has also been addressed \cite{1991Migli, 1993Kam, 1994Moraghe, 2008Marcucci, Schwengner01}.

Recently, the M1 excitation has been investigated in the framework based on the Skryme functionals \cite{Nesterenko01, Nesterenko02, Tselyaev01}, also extended to include tensor effects \cite{Cao2009}. It has been shown that the results for the spin-flip resonance obtained by using different parameterizations do not appear as convincing interpretation of the experimental results.
Additional effects have been explored in order to resolve this issue, e.g., the isovector-M1 response versus isospin-mixed responses, and the role of tensor and vector spin-orbit interactions~\cite{Nesterenko01, Nesterenko02}. 
In recent analysis in Ref.~\cite{Speth_2020}, based on the Skyrme functionals, it has been shown that while modern Skyrme parameterizations successfully reproduce electric excitations, there are difficulties to describe magnetic transitions. 
In addition, some Skyrme sets result in the ambiguity that, by the same parameterization, the model cannot simultaneously describe one-peak and two-peak data for closed and open shell nuclei~\cite{Nesterenko01}.
Thus, further developments of the Skyrme functional in the spin channel are called for~\cite{Speth_2020}.
Simultaneously, it is essential to explore the M1 response from different theoretical approaches to achieve a complete understanding of their properties, as well as to assess the respective systematic uncertainties. 

The aim of this work is to describe the properties of M1 excitations from a different perspective, 
by implementing a novel theory framework derived from the relativistic nuclear energy density 
functional. This framework has been in the past successfully employed in description of a variety of
nuclear properties and astrophysically relevant processes~\cite{Vretenar_01,2005Niksic,Paar2007,Paar2008,Niu2009,Khan2011,Paar2014,RocaMaza2015,Niksic2015,Paar2015,Vale2016,RocaMaza2018,Niksic01,Yuksel01}. In open shell nuclei, the pairing correlations make 
unnegligible contributions to the properties of M1 transitions~\cite{Pai01,2019OP}, 
thus they are also included in model calculations. 

The paper is organized as follows. In Sec. \ref{sec:FORMLSM} the overview of the formalism of the relativistic quasiparticle random phase approximation (QRPA) 
for magnetic transitions based on the relativistic point-coupling interaction is given. Section \ref{sec:SUMRULE_TEST} is devoted to display the benchmark result 
of our relativistic QRPA scheme in comparison with the sum rule in Ref. \cite{2019OP}. 
The results of model calculations and comparison with the experimental studies are presented in Secs. \ref{sec:Pb208RES} and \ref{sec:Ca48RES}, while the pairing effects are considered in Sec. \ref{sec:PAIRING}. A summary of the present work is given in Sec. \ref{sec:SUM}.

\section{\label{sec:FORMLSM} Formalism}
We study M1 excitations based on particle-hole ($1p \mhyphen 1h$) (or in open shell nuclei two-quasiparticle) transitions from 
$0^{+}$ ground state (GS) to $1^{+}$ excited states of even-even nuclei 
within the formalism of a relativistic nuclear energy density functional (RNEDF), 
assuming the spherical symmetry \cite{Niksic01}. 
More details about the RNEDF and its implementations
are given in Refs. \cite{Niksic01, Niksic03}. 
In this work, the nuclear ground state has been calculated 
by employing the self-consistent relativistic Hartree-Bogoliubov (RHB) model~\cite{Niksic01}, 
where the mean field is derived for the relativistic point-coupling interaction with density-dependent couplings. 
Many effects that go beyond the mean-field level are not explicitely included in the
RHB model, e.g., Fock terms, vacuum polarization effects and the short-range Brueckner-type correlations. Since the parameters of the RNEDF
are adjusted to the experimental data which contain all these and other effects, it means that effects beyond the mean-field level are implicitly included in the RHB
approach by adjusting the model parameters to reproduce a selected empirical data set~\cite{Vretenar_01}. The no-sea approximation is also employed for relativistic mean-field calculations~\cite{Vretenar_01}. 

Here we briefly present the formalism of the relativistic point coupling interaction starting from the Lagrangian density,
\begin {equation}\label{eq:1}
\begin{aligned}
 \mathcal{L} &=  \bar{\Psi}_{N} (i \gamma^{\mu}\partial_{\mu} - m_{0N}) \Psi_{N} \\
&- \frac{1}{2} \alpha_{S}(\rho)( \bar{\Psi}_{N}{\Psi}_{N} ) ( \bar{\Psi}_{N}{\Psi}_{N} )\\
& - \frac{1}{2} \alpha_{V}(\rho)( \bar{\Psi}_{N}\gamma^{\nu}{\Psi}_{N} ) ( \bar{\Psi}_{N}\gamma_{\nu}{\Psi}_{N} )\\
&- \frac{1}{2} \alpha_{TV}(\rho)( \bar{\Psi}_{N}\vec{\tau}\gamma^{\nu}{\Psi}_{N} )\cdot ( \bar{\Psi}_{N}\vec{\tau}\gamma_{\nu}{\Psi}_{N} ) \\
& -\frac{1}{2} \delta_{S} (\partial_{\nu}\bar{\Psi}_{N}{\Psi}_{N})(\partial^{\nu}\bar{\Psi}_{N}{\Psi}_{N}) \\
&- e \bar{\Psi}_{N} (\gamma^{\nu}A_{\nu}(\vec{x}))\frac{1 - \hat{\tau_{3}}}{2} \Psi_{N}.
\end{aligned}
\end {equation}
The first term corresponds to the free nucleon field of Dirac type, 
while the point coupling interaction terms include  isoscalar-scalar ($J^{\pi}=0^{+}$), 
isoscalar-vector ($J^{\pi}=1^{-}$), isovector-vector ($J^{\pi}=1^{-}$) channels (where $J$ denotes the respective quantum number for the angular momentum and $\pi$ for the parity), 
coupling of protons to the electromagnetic field, 
and the derivative term accounting for the leading effects of finite-range 
interactions necessary for a quantitative description of nuclear density distribution and radii.

The density dependent couplings in each channel, 
$\alpha_{S}(\rho)$ (isoscalar-scalar), 
$\alpha_{V}(\rho)$ (isoscalar-vector), and 
$\alpha_{TV}(\rho)$ (isovector-vector) are modeled 
by well-behaved functional~\cite{Niksic01}, 
\begin {equation}
\begin{aligned}
 \alpha_{i}(\rho) = a_{i} + (b_{i} + c_{i} x) e^{-d_{i}x}, 
\end{aligned}
\end {equation}
where $x = \frac{\rho}{\rho_{sat.}}$ and $\rho_{sat.}$ 
denotes nucleon density at saturation point for the case of 
symmetric nuclear matter. 
The respective parameters for each channel $i=S,V,TV$ are denoted 
as $a_{i}$, $b_{i}$, $c_{i}$ and $s_{i}$, while $\delta_{S}$ 
denotes the strength of isoscalar-scalar derivative term. 
In this work the DD-PC1 parameterization is used in model
calculations~\cite{Niksic01}.
The RHB model employed in this study includes pairing correlations 
described by the pairing part of the phenomenological Gogny interaction~\cite{Vretenar_01}, 
\begin{equation}
\begin{aligned}
 V^{pp}(1,2) = \sum_{i=1, 2} &e^{\lbrack (\bm{r}_{1} - \bm{r}_{2})/\mu_{i}\rbrack^{2}} \left( W_{i}+B_{i}\hat{P}^{\sigma}  \right. \\
  & \left. - H_{i}\hat{P}^{\tau} - M_{i}\hat{P}^{\sigma}\hat{P}^{\tau}  \right),
\end{aligned}
\end{equation}
where $\hat{P}^{\sigma}$ and $\hat{P}^{\tau}$ indicate the 
exchanges of the spin and isospin, respectively. 
The parameters $\mu_{i}$, $W_{i}$, $B_{i}$, $H_{i}$ and $M_{i}$ ($i = 1,2$) 
are given by the D1S set as in Ref. \cite{Berger}. 
We confirmed that, combined with the DD-PC1 functional for the mean-field part, 
this parameterization sufficiently reproduces the empirical 
pairing gaps of open-shell systems in this study.

For the M1 excitations of nuclei, we utilize the relativistic quasiparticle random phase approximation (RQRPA) \cite{Paar2003}, that is in this work developed for the implementation of the relativistic point-coupling interaction, extended to describe the unnatural-parity transitions of M1 type. 
In the limit of small amplitude oscillations, the RQRPA matrix equations read, 
\begin{equation}
\begin{pmatrix}
    A^{J}     &  B^{J} \\
    B^{*J}   &  A^{*J}  
\end{pmatrix} 
\begin{pmatrix}
    X^{\nu,JM} \\
    Y^{\nu, JM}   
\end{pmatrix} 
=
\hbar\omega_{\nu}
\begin{pmatrix}
    1   &  0 \\
    0   &  -1  
\end{pmatrix} 
\begin{pmatrix}
    X^{\nu,JM} \\
    Y^{\nu, JM}   
\end{pmatrix} 
\end{equation}
with $\hbar\omega_{\nu} = E_{\nu} - E_{0}$, 
where $E_{0}$ and $E_{\nu}$ are the RHB-ground and excitation energies of the 
many particle system, respectively. 
The $X^{\nu}$ and $Y^{\nu}$ indicate the forward-scattering and backward-scattering 
two-quasiparticle amplitudes. 
The particle-hole channel of the residual RQRPA interaction, $V^{ph}$, 
has been calculated by the effective Lagrangian from Eq. (\ref{eq:1}), 
but supplemented with the isovector-pseudovector interaction, as we 
explain in the following section. 
The RQRPA particle-particle correlations, $V^{pp}$, is evaluated from the 
Gogny-D1S force, which is commonly used in the RHB model. 

The RHB calculations in this work are
performed in the computational framework developed in Refs. \cite{Niksic01, Paar2003, Niksic02, Niksic03}. 
The ground state of spherical nucleus is solved in the 
model space expanded in the harmonic oscillator (HO) basis, 
including up to 20 shells. The cutoff energies for the configuration
space in the QRPA are selected to provide a sufficient convergence 
in the M1-excitation strength. 

\subsection{Isovector-pseudovector interaction}
In order to describe the unnatural parity excitations of the M1 type ($J^{\pi}=1^{+}$), 
the RQRPA residual interaction is further extended by introducing the 
relativistic isovector-pseudovector (IV-PV) contact interaction, 
\begin {equation}
\begin{aligned}
\mathcal{L}_{IV-PV} = -\frac{1}{2}\alpha_{IV-PV} \lbrack  \bar{\Psi}_{N} \gamma^{5} \gamma^{\mu} \vec{\tau}  \Psi_{N} \rbrack \cdot \lbrack  \bar{\Psi}_{N} \gamma^{5}  \gamma_{\mu} \vec{\tau} \Psi_{N}  \rbrack.
\end{aligned}
\end {equation}
This pseudovector type of interaction has been modeled as a scalar product of two pseudovectors. 
The strength parameter for this channel, $\alpha_{IV-PV}$, is considered 
as a parameter, which is constrained by the experimental data on 
M1 transitions of selected nuclei. 
We note that the IV-PV term does not contribute in the RHB calculation of 
the ground state, thus its strength parameter cannot be constrained together 
with other model parameters on the bulk properties of nuclear ground state.
The pseudovector type of interaction would lead to the parity-violating mean-field at the Hartree level for the description of 
the $\text{0}^+$ nuclear ground state, and it contributes only to the RQRPA equations 
for unnatural parity transitions, i.e. $1^+$ excitation of M1 type.

The coupling strength parameter $\alpha_{IV-PV}$ is determined by minimizing the standard deviation $\sigma_{\Delta}(\alpha_{IV-PV})$, where  $\Delta$ is the gap between the theoretically calculated centroid energy and experimentally determined dominant peak position of measured M1 transition strength in  $^{208} \rm Pb$ \cite{Laszewski} and $^{48} \rm Ca$  \cite{2016Birkhan} nuclei. 
It turns out that the optimal parameter value is $\alpha_{IV-PV} =  0.53\ \rm MeV fm^{3}$  and in this case the energy gap $\Delta$ is less than 1 MeV both for $^{208} \rm Pb$ and $^{48} \rm Ca$. In this way all the parameters employed in the present analysis are constrained and further employed in the analysis of the properties of M1 excitations.

\subsection{\label{sec:ISOVS}  Transition strength for M1 excitations}

In the following we give an overview of the formalism for the transition
strength for M1 transitions, for the implementation in the RQRPA. 
The transition strength for magnetic multipole excitations $B(MJ,\omega_{\nu})$ can be
distinguished to isoscalar strength  $B^{(IS)}(MJ,\omega_{\nu})$, isovector strength $B^{(IV)}(MJ,\omega_{\nu})$, as well as spin $B^{\sigma}(MJ,\omega_{\nu})$ and orbital $B^{\ell}(MJ,\omega_{\nu})$ strengths. From the transition strength distribution of interest, the energy weighted moment $m_{k}$ and centroid energy $\bar{E}$ can be calculated.

Within the (Q)RPA framework, discrete spectrum $B(MJ,\omega_{\nu})$ of excited states is obtained. 
For demonstration purposes, this quantity is convoluted with the Lorentzian distribution~\cite{Paar2003},
\begin{equation}\label{eq:5}
\begin{aligned}
R_{MJ}(E) = \sum_{\nu} B(MJ,\omega_{\nu}) \frac{1}{\pi} \frac{\Gamma/2}{ (E - \hbar\omega_{\nu})^{2} + (\Gamma/2)^{2}  },
\end{aligned}
\end{equation}
where the Lorentzian width is set as $\Gamma = 1.0 \rm \ MeV$. 
The discrete strength $B(MJ,\omega_{\nu})$ for the magnetic operator $\hat{\mu}_{JM}$ of rank $J$ is, 
within the spherical assumption, calculated by the following expression~\cite{Paar2003},
\begin{equation}
\label{BM1expr}
\begin{aligned}
B(MJ,\omega_{\nu}) = &\Big\vert \sum_{\kappa \kappa'} \Big( X^{\nu,J 0}_{\kappa \kappa'} \langle \kappa||\hat{\mu}_{J} ||\kappa' \rangle
\\&
+
(-1)^{j_{\kappa} - j_{\kappa'} + J}Y^{\nu,J 0}_{\kappa \kappa'} \langle \kappa'||\hat{\mu}_{J} ||\kappa \rangle
      \Big) 
\\
&\times \Big(u_{\kappa}v_{\kappa'} + (-1)^{J}v_{\kappa}u_{\kappa'} \Big)
\Big\vert^{2},
\end{aligned}
\end{equation}
where $\kappa$ and $\kappa'$ are quantum numbers which are denoting single-particle states in the canonical basis~\cite{Paar2003}. 
The $u_{\kappa}$ and $v_{\kappa}$ are the RHB occupation coefficients of single-particle states.

In the case of M1 excitations ($0^{+} \rightarrow 1^{+}$), the rank of the transition operator is $J = 1$. 
The reduced matrix element
for the M1 operator $\hat{\bm{\mu}}$, in mixed spin-isospin basis, 
is given by,
\begin{equation}
\begin{aligned}
\langle j_{f}; t_{f}, t_{zf}|| \hat{\bm{\mu}}|| j_{i}; t_{i}, t_{zi} \rangle = 
\Big(  \langle j_{f} || \hat{\bm{\mu}}^{(IS)} || j_{i}\rangle   
\langle t_{f} || 1_{\tau} || t_{i} \rangle 
\\  - 
\frac{C_{M1}}{\sqrt{2t_{f} + 1}} \langle j_{f} ||\hat{\bm{\mu}}^{(IV)} || j_{i} \rangle 
\langle t_{f}  || \hat{\vec{\tau}} || t_{i}\rangle   \Big),
\label{rme1}
\end{aligned}
\end{equation}
where the two resulting matrix elements correspond to the transitions of isoscalar and isovector type. 
Here $C_{M1} = \langle \frac{1}{2} t_{z}; 10   | \frac{1}{2} t_{z}  \rangle $ 
denotes Clebsch-Gordan coefficient in isospin space  with 
convention $t_{z} = \frac{1}{2}$ for neutrons 
($C_{M1} = \frac{1}{\sqrt{3}}$) and $t_{z} = - \frac{1}{2}$ for 
protons ($C_{M1} = - \frac{1}{\sqrt{3}}$). 

A complete expression of M1 operator in the relativistic formalism which acts on Hilbert space with mixed spin-isospin basis is given in a block diagonal form,
\begin{equation}
\label{M1operator}
\begin{aligned}
\hat{\bm{\mu}}_{1\nu} = 
\sum_{k = 1}^{A}
\begin{pmatrix} 
 \hat{\bm{\mu}}_{1\nu}^{(IS)}(11)_{k}   & 0 \\
 0 & \hat{\bm{\mu}}_{1\nu}^{(IS)}(22)_{k}
\end{pmatrix}  
\otimes 1_{\tau} \\
 -  \sum_{k = 1}^{A} 
\begin{pmatrix} 
 \hat{\bm{\mu}}_{1\nu}^{(IV)}(11)_{k}  & 0 \\
 0 & \hat{\bm{\mu}}_{1\nu}^{(IV)}(22)_{k}
\end{pmatrix}  
\otimes \hat{\tau}_{3},
\end{aligned}
\end{equation}
where $1_{\tau}$ and  $\hat{\tau}_{3}$ are unit and Pauli's $2\times2$ matrices in isospin space. 
The $\hat{\bm{\mu}}_{1\nu}^{(IS)}$ and $\hat{\bm{\mu}}_{1\nu}^{(IV)}$  components 
correspond to the isoscalar and isovector part of the M1 operator for $k^{th}$ 
nucleon, given by
\begin{equation}
\label{operatorISIV}
\begin{aligned}
&\hat{\bm{\mu}}_{1\nu}^{(IS,IV)}(11)_{k} =  \hat{\bm{\mu}}_{1\nu}^{(IS,IV)}(22)_{k}
\\
&= \frac{\mu_{N}}{\hbar} (g^{IS,IV}_{\ell}\hat{\bm{\ell}}_{k} +  g^{IS,IV}_{s}\hat{\bm{s}}_{k}) \cdot \bm{\nabla} ( r Y_{1\nu}(\Omega_{k}) ). 
\end{aligned}
\end{equation}
The empirical gyromagnetic ratios for the bare proton ($\pi$) and neutron ($\nu$) are given as 
$g^{\pi(\nu)}_{\ell}=1~(0)$ and $g^{\pi(\nu)}_{s}=5.586~(-3.826)$~\cite{Fujita}, 
where the units are given in nuclear magneton, $\mu_{N} = e\hbar/(2m_{\rm N})$. 

In this work, so-called isoscalar and isovector gyromagnetic ratios are determined separately for the orbital and spin components \cite{Lipparini1976, Fujita2011}.
That is,
\begin{equation}
\label{gIS}
\begin{aligned}
g^{IS}_{\ell} &= \frac{g^{\pi}_{\ell} + g^{\nu}_{\ell}}{2} = 0.5, 
&
g^{IS}_{s} &= \frac{g^{\pi}_{s} + g^{\nu}_{s}}{2} = 0.880, 
\end{aligned}
\end{equation}
and
\begin{equation}
\label{gIV}
\begin{aligned}
g^{IV}_{\ell} &= \frac{g^{\pi}_{\ell} - g^{\nu}_{\ell}}{2} = 0.5,
& 
g^{IV}_{s} &= \frac{g^{\pi}_{s} - g^{\nu}_{s}}{2} = 4.706.  
\end{aligned}
\end{equation}
This decomposition of the isoscalar and isovector M1 operators is consistent with the non-relativistic formalism already used in previous studies, e.g. Refs. \cite{Lipparini1976, Fujita2011}. More details 
are given in Appendix \label{app:ISIV_DECO}.

We notify that in the probe-independent consideration of nuclear excitations,
the IS and IV operators have equal weights.
In this manner the IS or IV character is a structure feature of the nucleus, and it is independent of the probe, which can be either strong, electromagnetic, or weak~\cite{Sagawa2019}. In the present M1 case, the $g_s$ factors in the IS and IV operators are different, and are probe-dependent \cite{Sagawa2019, Love1983}.
Convention given in Eqs. (\ref{gIS}) and (\ref{gIV}) corresponds to the electromagnetic process~\cite{Sagawa2019}.
As pointed out in Ref.~\cite{Sagawa2019}, for the electromagnetic processes, the IS-spin g-factor is considerably smaller than the IV one, i.e. $(g^{IS}_s/g^{IV}_s)^2\approx 1/30$, as one can see from the values given above.
Note also that, for the hadronic processes, the IS-spin coupling is still smaller than the IV one, but with an enhancement at intermediate energy~\cite{Love1983, Sagawa2019}.

Previous studies also addressed possible quenching effects on the M1 mode. 
Several theoretical descriptions of the total M1 transition strength result in the over-estimated values, in comparison to the experiments (see review in Ref.~\cite{Richter03}). One phenomenological solution is to introduce the in-medium effects by quenching the gyromagnetic factors, $g_{s,~\ell}$ \cite{Richter03, Nesterenko01, Nesterenko02}. 
In addition, experiments report the fragmented strength, which is, from theoretical point of view, a result of the couplings involving complex configurations, 
e.g., of two-particle-two-hole $(2p \mhyphen 2h)$ interactions as pointed out, for example, in Ref. \cite{Dehesa}. 
However, introduction of $2p \mhyphen 2h$ effects is technically demanding task going beyond the scope of this work, and we leave it for the future study. 
In this work, we use the $g$-factors of the bare nucleons, i.e. the quenching effect has not been considered. 

The reduced matrix elements for
the isoscalar or isovector component of M1 operator in Eq. (\ref{M1operator}) is given by,
\begin{equation}
\begin{aligned}
\langle j_{f}||\hat{\bm{\mu}}^{(X)}|| j_{i}\rangle &= I_{large} \langle (\frac{1}{2},\ell_{f}) j_{f}|| \hat{\bm{\mu}}^{(X)}(11)|| (\frac{1}{2},\ell_{i}) j_{i}\rangle
\\
&+I_{small} \langle(\frac{1}{2},\tilde{\ell}_{f}) j_{f}|| \hat{\bm{\mu}}^{(X)}(22)|| (\frac{1}{2},\tilde{\ell}_{i}) j_{i}\rangle,
\end{aligned}
\end{equation}
where $X=IS$ or $IV$, respectively. 
Here the radial parts of these matrix elements are given by the integrals,
\begin{equation}
\begin{aligned}
I_{large} =    \int_{0}^{\infty} f^{*}_{n_{f}j_{f}}(r)  f_{n_{i}j_{i}}(r)  r^{2}dr ,
\end{aligned}
\end{equation}
and 
\begin{equation}
\begin{aligned}
I_{small} = \int_{0}^{\infty} g^{*}_{n_{f}j_{f}}(r) g_{n_{i}j_{i}}(r)   r^{2}dr,
\end{aligned}
\end{equation}
where $f_{nj}(r)$ are large and $g_{nj}(r)$ small radial components of the quasiparticle Dirac spinors for nucleons~\cite{Niksic01}.
The labels $j_{i}$ and $j_{f}$ denote total nucleon
angular momenta of the initial and final state, respectively. The orbital angular momenta that correspond to large ($l_i$,$l_f$) and small $(\tilde{l_f},\tilde{l_i})$ spinor components are determined by the total angular momenta $(j_{i},j_{f})$ and parity $({\pi}_{i},{\pi}_{f})$ of the initial and final states~\cite{Niksic01}. 
Thus, the M1 transition strength in the RQRPA formalism is given by,
\begin{equation}\label{eq:10}
\begin{aligned}
B(M1,E) = \Big\vert \sum_{\kappa \kappa'} \Big( X^{\nu,1 0}_{\kappa \kappa'} -(-1)^{j_{\kappa} - j_{\kappa'}}Y^{\nu,1 0}_{\kappa \kappa'} \Big) 
\\
\times \Big(u_{\kappa}v_{\kappa'} - v_{\kappa}u_{\kappa'} \Big) 
\langle j_{\kappa'}; \frac{1}{2} \ t_{z} || \hat{\bm{\mu}}|| j_{\kappa}; \frac{1}{2} \ t_{z} \rangle 
\Big\vert^{2}.
\end{aligned}
\end{equation}
Similarly, following Eq.~(\ref{rme1}), the isovector transition strength reads
\begin{equation}\label{eq:15}
\begin{aligned}
B^{(IV)}(M1,E) =  \Big\vert   -  \sum_{\kappa \kappa'} \frac{C_{M1}}{\sqrt{2t_{f} + 1}} \langle \frac{1}{2}  || \hat{\vec{\tau}} || \frac{1}{2}\rangle\\
\times  \Big( X^{\nu,1 0}_{\kappa \kappa'}  -(-)^{j_{\kappa} - j_{\kappa'}}Y^{\nu,1 0}_{\kappa \kappa'} \Big) 
\\
\times \Big(u_{\kappa}v_{\kappa'} - v_{\kappa}u_{\kappa'} \Big) 
\langle j_{\kappa'} ||\hat{\bm{\mu}}^{(IV)} || j_{\kappa} \rangle  \Big\vert^{2},
\end{aligned}
\end{equation}
whereas the isoscalar strength is given by,
\begin{equation}\label{eq:20}
\begin{aligned}
B^{(IS)}(M1,E) &= \Big\vert  \sum_{\kappa \kappa'} \langle \frac{1}{2} || 1_{\tau} || \frac{1}{2} \rangle  \Big( X^{\nu,1 0}_{\kappa \kappa'} 
-
(-)^{j_{\kappa} - j_{\kappa'}}Y^{\nu,1 0}_{\kappa \kappa'} \Big) 
\\
\times & \Big(u_{\kappa}v_{\kappa'} - v_{\kappa}u_{\kappa'} \Big) 
\langle j_{\kappa'} || \hat{\bm{\mu}}^{(IS)} || j_{\kappa}\rangle  \Big\vert^{2}.
\end{aligned}
\end{equation}
In the following calculations, we also refer to 
the spin-M1 transition strength, 
\begin{equation}
\begin{aligned}
  B^{\sigma}(M1,E) = B(M1,E)\vert_{g_{\ell} = 0}, \label{eq:YDRQ_s}
\end{aligned}
\end{equation}
where the orbital gyromagnetic factors are set to zero. 
Similarly, the orbital M1 strength is given as 
\begin{equation}
\begin{aligned}
  B^{\ell}(M1,E) = B(M1,E)\vert_{g_{s} = 0}. \label{eq:YDRQ_l}
\end{aligned}
\end{equation}
For the analysis of M1 transition strength, energy-weighted moments $m_{k}$ 
of the discrete spectra are used: 
\begin{equation}
\label{eq:moments}
m_{k} = \sum_{\nu} B(MJ,\omega_{\nu}) (E_{\nu} - E_{0} )^{k}.
\end{equation}
Two moments are often used in order to compare experimental results with theoretical predictions, i.e. 
the non-energy-weighted sum $m_{0}$, which corresponds to the total strength $B(M1)$, 
and energy-weighted sum of the strength $m_{1}$. 
A quotient of two moments, $\bar{E} = m_{1}/m_{0}$, corresponds to the
centroid energy which represents an average energy of discrete strength distribution. 

\subsection{The M1 sum rule in core-plus-two-nucleon systems} \label{sec:SUMRULE_TEST}

For the consistency check of numerical calculations of nuclear excitations, such as giant resonances, 
the sum rules related to the transition strength have provided in the past
a useful guidance \cite{Speth01, Suzuki01, Liu01, Oesterfeld01, Weisskopf01, Ring01}. 
In this section, the M1 sum rule introduced in Ref. \cite{2019OP} is utilized to test the validity of the framework established in this work.  
In Ref. \cite{2019OP}, the non-energy weighted sum ($m_{k=0}$) of the M1 
excitation was evaluated for some specific systems, which 
consist of the core with shell-closure and additional 
two valence neutrons or protons, e.g., $^{18}$O and $^{42}$Ca. 
If the pairing correlations between the valence nucleons 
are neglected, one advantage of that sum rule is that 
its non-energy weighted sum-rule value (SRV) is determined analytically 
for the corresponding system of interest. 

\begin{figure}[t] \begin{center}
  \includegraphics[width = 0.95\hsize]{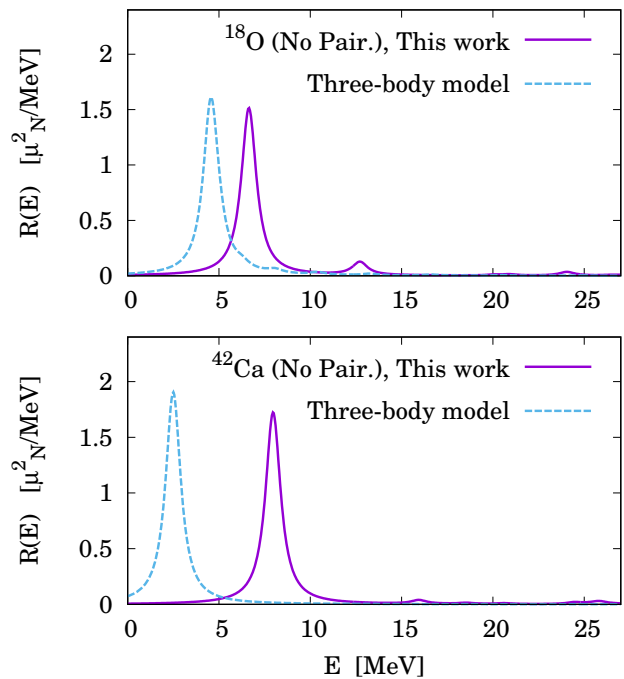}
  \caption{(Top) The M1 transition strength distributions for $^{18}$O (upper panel)
  and $^{42}$Ca (lower panel) based on the RRPA calculations with DD-PC1 parameterization.
The result from the three-body model without pairing from 
Ref. \cite{2019OP} is shown for comparison. 
} \label{fig:1221_2018}
\end{center} \end{figure}
\begin{table}[b]
\begin{center}
\caption{Non-energy-weighted sum ($m_{k=0}$) of the M1-response function for $^{18}$O and $^{42}$Ca 
obtained from the RRPA calculation. 
The analytical SRV value from Ref. \cite{2019OP} is shown for comparison. 
The unit is $\mu_{\rm N}^2$. }   \label{table:app_1_1}
  \begingroup \renewcommand{\arraystretch}{1.2}
  \begin{tabular*}{\hsize} { @{\extracolsep{\fill}} lcc } \hline   \hline
                      & $m_{0}$ (This work)    & SRV \cite{2019OP}   \\ \hline
  ~$^{18}$O    &$2.73$          &$2.79$ \\
  ~$^{42}$Ca   &$2.91$          &$2.99$ \\  \hline  \hline
  \end{tabular*}
  \endgroup
\end{center}
\end{table}

In this section, we perform the calculation of M1 SRV within the relativistic random phase approximation (RRPA), based on 
the formalism given in Sec.~\ref{sec:FORMLSM}.  
The DD-PC1 functional is used, supplemented with the IV-PV channel 
in the RRPA residual interaction, with the strength 
parameter $\alpha_{IV-PV}=0.53$ MeV$\cdot{\rm fm}^3$, as introduced in Sec.~\ref{sec:FORMLSM}. 
Note that in this sum rule test,  the pairing correlations are neglected in calculations. 
Figure \ref{fig:1221_2018} displays the M1-response function $R(E)$ from
Eq. (\ref{eq:5}) for $^{18}$O and $^{42}$Ca, obtained with the RRPA. 
Our result shows the dominant single peak in each system. 
Table \ref{table:app_1_1} shows the non-energy-weighted sum ($m_0$) results 
for M1 transitions in $^{18}$O and $^{42}$Ca obtained from the RRPA. 
The $m_0$ value is calculated from Eq. (\ref{eq:moments}), using the M1-strength 
distribution up to $50$ MeV. 
For comparison, the respective values introduced
in Ref.~\cite{2019OP} are also shown as ``SRV''. 
The RRPA accurately reproduces the SRVs for two nuclei under consideration. Namely, the relativistic framework to describe the M1 
transitions appears to be valid at the level of no-pairing limit. 
The small deficiency of RRPA beyond the SRV value is attributable 
to the cutoff energy. 

As shown in Fig.  \ref{fig:1221_2018},
the RRPA excitation energy of M1 mode appears rather different than in the
case of three-body model from Ref. \cite{2019OP}. This discrepancy originates
from different open-shell structures produced by the two models. 
For further improvement, 
one may adjust these models directly to the M1-reference data, which have been, 
however, not precisely obtained for $^{18}$O neither $^{42}$Ca. 
Nevertheless, since the sum rule does not depend on the excitation energy, 
the analysis of both approaches confirms the expected M1-sum values, 
and thus, justifies our RRPA implementation. 

\section{Results}
In the following we present the results on M1 transitions based on the RHB+R(Q)RPA 
framework introduced in Sec. \ref{sec:FORMLSM}. 
The functional DD-PC1 \cite{Niksic01} is systematically used in model calculations, 
supplemented with the pairing interaction from the phenomenological Gogny-D1S force \cite{Berger}. 

\subsection{\label{sec:Pb208RES} M1 transitions in  $^{208} \rm Pb$ }
\begin{figure}[t]
\includegraphics[scale=0.33]{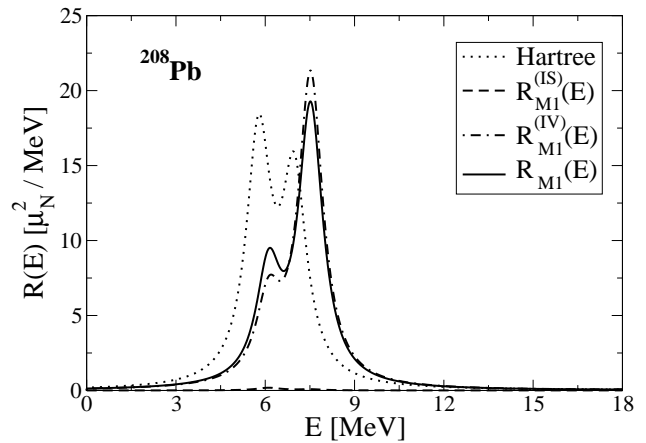}
\captionof{figure}{The M1 transition strength distribution for $^{208}$Pb.
The results are obtained with the RRPA using the DD-PC1 functional: 
$R_{M1}(E)$ for full, 
$R^{(IS)}_{M1}(E)$ for isoscalar, and 
$R^{(IV)}_{M1}(E)$ for isovector response functions. 
Unperturbed response at the Hartree level is also shown. }
\label{fig:Pb01}
\end{figure}

As the first case for detailed analysis of M1 transitions in the framework
based on the RNEDF, we the consider $^{208} \rm Pb$ nucleus. There is experimental data on 
this system available \cite{Raman02, Laszewski04, Raman01, Laszewski02, Laszewski, 2016Birkhan}, 
and thus, it is suitable to the first application.
Figure \ref{fig:Pb01} shows the M1-response function $R_{M1}(E)$ for $^{208} \rm Pb$, obtained using Eq.(\ref{eq:5}). 
Since it is magic nucleus, pairing correlations do not contribute to the nuclear ground state energy, 
and the RHB + RQRPA reduces to 
the relativistic Hartree + RRPA model. In addition to the full response $R_{M1}(E)$, 
the responses to the isoscalar and isovector operators, 
$R_{M1}^{IS} (E)$ and $R_{M1}^{IV} (E)$, are shown separately. 
For comparison, the so-called ``unperturbed'' response at the Hartree level is also shown, corresponding to the limit when the residual RRPA interaction is set to zero.

As shown in Fig. \ref{fig:Pb01}, the full M1 response is dominated by the two peaks at $6.11$ MeV and $7.51$ MeV. 
These two peaks exhaust most of the total M1 strength up to $50$ MeV energy. 
In some experimental studies \cite{Laszewski, 2016Birkhan}, there has been 
measured a dominant peak of the M1-strength distribution 
around $7.0$-$7.5$ MeV, whereas the other bump could be found at $\cong 6.2$ MeV \cite{Laszewski}. 
However, the experimental data also show a more fine fragmentation of the M1 strength in $^{208}$Pb \cite{Laszewski, 2016Birkhan}. 
To reproduce this fragmentation, the present RRPA may need to be improved with, 
e.g. the two-particle-two-hole effect \cite{Dehesa}, which is beyond the present scope. 
Even with a lack of details,
our  RRPA scheme reproduces the rough structure of the M1-distribution of $^{208}$Pb, especially its dominant two-peak structure. 
By comparing the full-M1 response with 
the unperturbed response at the Hartree level (Fig. \ref{fig:Pb01}), 
one can observe that the full response is shifted to higher energies, 
demonstrating the effect of the IV-PV residual interaction to establish 
M1 transitions as genuine nuclear mode of excitation. 
This shift is consistent to the selection rule of M1, since 
the IV-PV interaction affects the $J^{\pi}=1^+$ excited states. 

In Fig. \ref{fig:Pb01}, one can observe that the isovector M1 response is significantly larger than the isoscalar one, and both components interfere.
The dominance of isovector mode is visible also from the M1 strength integrated up to 50 MeV:
the full strength amounts $\sum B(M1) = 41.99\ \mu^{2}_{N}$, whereas 
the isoscalar strength is $\sum B^{(IS)}(M1) = 0.43\ \mu^{2}_{N}$ and 
the isovector strength is $\sum B^{(IV)}(M1) = 42.33\ \mu^{2}_{N}$. 
The main reason of the isovector-M1 dominance is a large difference 
in the respective gyromagnetic ratios, 
$g_s^{IS}$ and $g_s^{IV}$ given in Eqs. (\ref{gIS}) and (\ref{gIV}).
For more details, see Appendix \ref{app:ISIV_DECO}.

\renewcommand{\arraystretch}{1.5}
\begin{table}[b]
\begin{center}
  \caption{Partial neutron and proton contributions $(b_{ph}^{\nu,\pi})$ to the M1 transition strength for the two main peaks $(E_{peak})$, and respective overall transition strengths B(M1) for $^{208}$Pb. Details about configurations are given in the text.}  \label{table:Th_Pb208_04}  
   \begin{tabular*}{\hsize} { @{\extracolsep{\fill}} cccc}
   \hline  \hline
    $E^{th.}_{peak}  \lbrack \rm MeV \rbrack$   & $b_{ph}^{\nu} ~\lbrack \mu_{N} \rbrack$ &  $b_{ph}^{\pi} ~\lbrack \mu_{N} \rbrack$ &  $ B(M1) ~\lbrack \mu^{2}_{N} \rbrack$ \\  \hline  
      6.11  &  $-$1.33 &  4.74  & 11.6 \\
      7.51  &  4.22 &  1.15 & 28.96 \\
   \hline  \hline
   \end{tabular*}
\end{center}
\end{table}
\renewcommand{\arraystretch}{1}

The structure of the two pronounced M1 peaks is analyzed in more details. 
Table \ref{table:Th_Pb208_04} shows the respective partial contributions $b_{ph}$ to the $B({\rm M1})$ 
strength from the major proton and neutron particle-hole $(ph)$ configurations (see Eq. (\ref{BM1expr})).
That is, if a single proton $(p_1 h_1)$ and neutron $(p_2 h_2)$ configurations contribute to the M1 transition at excitation
energy $E$,
\begin{equation}
  B({\rm M1},~E) = \left| b^{\nu}_{p_1 h_1}(E) +b^{\pi}_{p_2 h_2}(E)  \right|^2. 
\end{equation}
For the state at $6.11$ MeV, the major contribution 
comes from the transitions between spin-orbit partner states for neutrons, 
$(\nu 1i^{-1}_{13/2}\rightarrow \nu1i_{11/2})$ and protons, $(\pi 1h^{-1}_{11/2}\rightarrow \pi 1h_{9/2} )$. 
As shown in Table \ref{table:Th_Pb208_04}, 
partial proton and partial neutron spin-flip transitions interfere destructively. 
In the case of the state at $7.51$ MeV, the main transitions are $(\nu 1i^{-1}_{13/2}\rightarrow \nu 1i_{11/2})$
and $(\pi 1h^{-1}_{11/2}  \rightarrow \pi 1h_{9/2} )$, with coherent contributions 
to the total B(M1) strength, as shown in Table \ref{table:Th_Pb208_04}. 

\begin{figure}[t]
\centering
\includegraphics[scale=0.33]{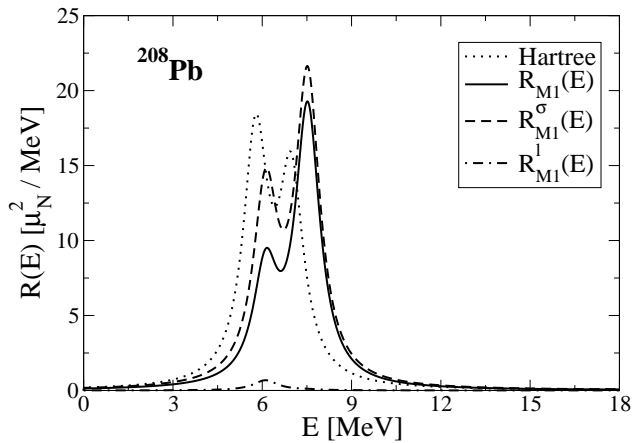}
\captionof{figure}{The same as Fig.~\ref{fig:Pb01}, but with R(M1) - full, $R_{M1}^{\sigma}(E)$ - spin and $R^{\ell}_{M1}(E)$ - orbital response functions. Unperturbed response at the Hartree level is also shown.
}
\label{fig:Pb02}
\end{figure}

Figure \ref{fig:Pb02} shows the separation of the full M1 response in $^{208}$Pb to
the spin and orbital response functions, $R^{\sigma}_{M1}(E)$ and $R^{\ell}_{M1}(E)$,
evaluated by Eqs. (\ref{eq:YDRQ_s}) and (\ref{eq:YDRQ_l}), respectively.
One can observe considerably larger spin response in 
comparison to the orbital one, i.e. most of the overall $B(M1)$ strength is exhausted by the spin M1 response. 
The orbital-M1 strength is mainly related to deformation and almost disappears in the closed-shell nuclei~\cite{Richter01}.
The spin and orbital responses interfere destructively, 
i.e. the full response function is smaller than the spin response. 
The corresponding sum of the strengths amount $53.14~\mu^{2}_{N}$ for the spin, 
$1.19~\mu^{2}_{N}$ for the orbital, and 
$41.99~\mu^{2}_{N}$ for the full M1 transition. 
In the present study of $^{208}$Pb, the overall M1-excitation strength is 
almost fully exhausted by the two peaks, i.e., 
$B(M1, E=6.11 \: {\rm MeV}) = 11.6 \ \mu^{2}_{N}$ and $B(M1, E=7.51 \: {\rm MeV}) = 28.96\ \mu^{2}_{N}$.

Over the past years, there have been several experimental studies of M1 transitions in $^{208}$Pb. 
Several experimental results are summarized 
in a chronological order in Table \ref{table:Exp_Pb208_01}. 
It turns out that the total measured strength lies between 
$\sum B(M1) \cong (35.0 - 47.5)\ \mu^{2}_{N}$ and it is in agreement with our result. 
However, from recent experiments \cite{Laszewski02, Laszewski, 2016Birkhan}, 
it has also been pointed out that the actual M1 sum could need to be 
reduced, mainly due to the confusion with the electric-dipole components 
in the old experiments. 
A recent theoretical investigation based on the Skyrme functional \cite{Tselyaev01} results 
even in lower $B(M1)$ strength in comparison to the modern experiments. 
Therefore, quantitative description of the transition strengths for M1 modes remains an open question.
In comparison to Bohr and Mottelson's independent particle model (IPM), 
which estimates $B(M1) = 36\ \mu^{2}_{N}$  in Ref. \cite{Laszewski03}, 
the RRPA result is comparable, but somewhat higher. In Refs. \cite{Nesterenko01, Nesterenko02, Tselyaev01}, the similar calculation based on the Skyrme functionals have been used, with the quenching factors in $g$ coefficients, $\simeq$ 0.65, in order to reduce the transition strength. When using the quenching, the total B(M1) strength for $^{208}$Pb obtained for a set of Skyrme parameterizations amounts 14.8-17.3 $\mu^{2}_{N}$~\cite{Nesterenko01}.

\renewcommand{\arraystretch}{1.4}
\begin{table}[t]
\begin{center}
\caption{A summary of reported experimental M1 excitation energies and transition strengths in $^{208}\rm Pb$. 
In Ref.~\cite{Raman01}, the symbol (*) denotes experimental ambiguity 
due to parity assignment to the quantum mechanical state with respect to M1 transitions.}
\label{table:Exp_Pb208_01}
   \begin{tabular*}{\hsize} { @{\extracolsep{\fill}} ccc}  \hline  \hline
  ($^{208}$Pb)  &$E_{x} \lbrack \rm MeV \rbrack$   & $B_{M1,~\uparrow} \lbrack \mu^{2}_{N}\rbrack$ \\ \hline
  Ref. \cite{Raman02} (1977) &(sum)  &$35.0$ \\
  \hline
  Ref. \cite{Laszewski04} (1977) &(sum)  &$46.5$  \\
  \hline
  Ref.~\cite{Raman01} (1979) &$4.843$      &$5.8^*$  \\
  &$7.061$      &  $17.7^*$ \\
  &$7.249$      &  $0.5$  \\
  &$7.37-7.82$  &  $7.9$  \\
  &$7.98$       &  $7.1^*$ \\
  &$8.20-9.50$  &  $8.5$  \\
  &$4.843 - 9.50$        &$\sum B(M1,E)_{\uparrow}$ \\
  &&  $=47.5~(30.6^*)$ \\
  \hline
  Ref. \cite{Laszewski02} (1985) &$5.8 - 7.4$ &  $10.7$  \\
  &$\leq 6.4$  & $1.9$  \\
  &7.3    & $15.6$ \\
 \hline
  Ref. \cite{Laszewski} (1988) &$6.7 - 8.1$ &  $19.0$ \\
 \hline
  Ref. \cite{2016Birkhan} (2016) &$7.0 - 9.0$  &$20.5$ \\
\hline  \hline
 \end{tabular*}
  \end{center}
\end{table}
\renewcommand{\arraystretch}{1}

\subsection{\label{sec:Ca48RES} M1 transitions in $^{48} \rm Ca$}
\begin{figure}[t]
  \includegraphics[scale=0.33]{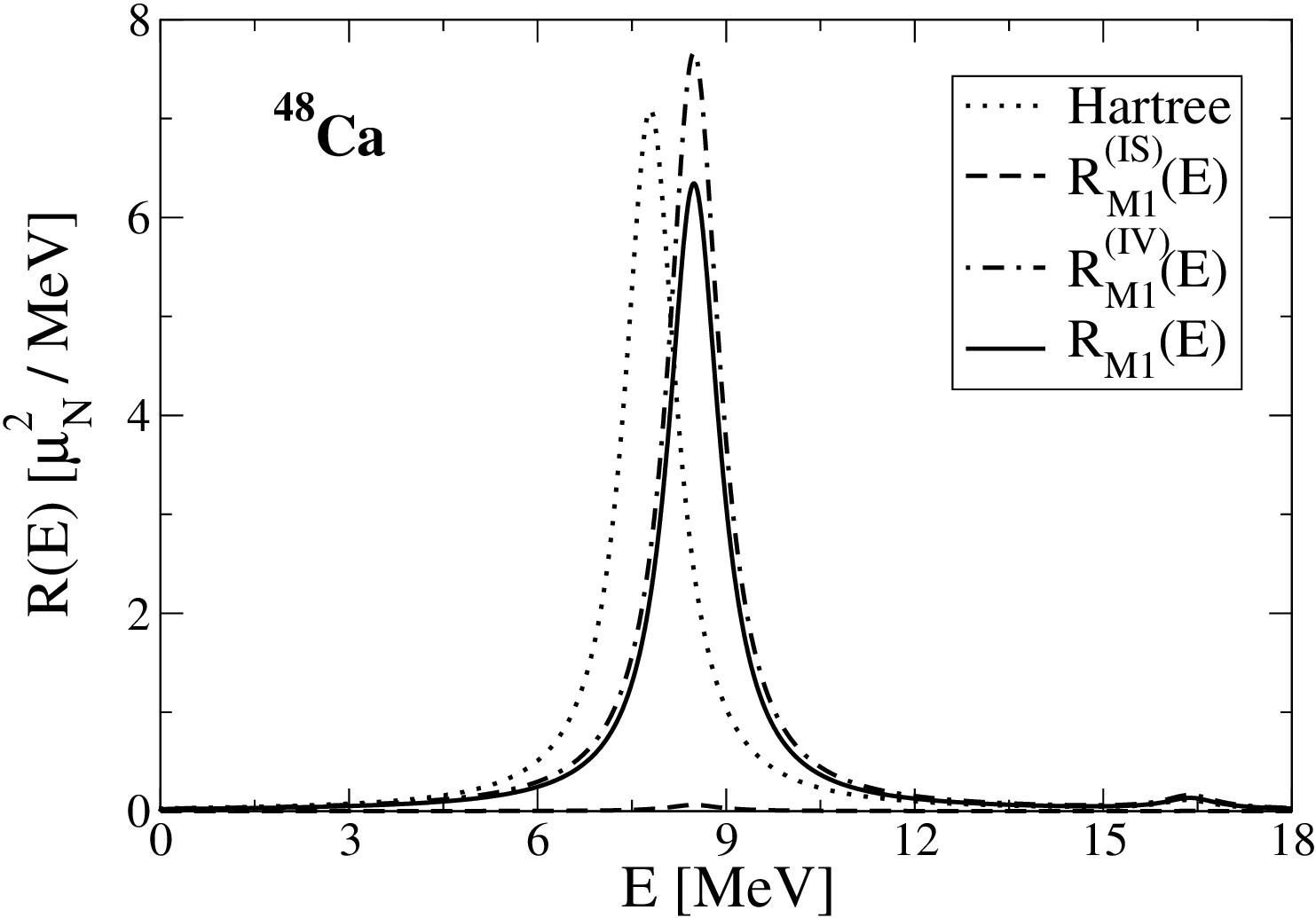}
  \captionof{figure}{The same as Fig. \ref{fig:Pb01}, but for $^{48} \rm Ca$.}
  \label{fig:Ca01}
\end{figure}

In the following, we extend our study to the lighter system, $^{48} \rm Ca$ nucleus, 
where several experimental data are available \cite{2016Birkhan, Yako2009, 2011Tompkins}. 
In Fig. \ref{fig:Ca01}, the RRPA full, isoscalar, and isovector $B(M1)$ transition-strength distributions are shown for $^{48} \rm Ca$. The M1 strength distribution is composed
from a single dominant peak.
The corresponding transition strength summed up to $50$ MeV 
amounts $B(M1)$ = 10.38 $\ \mu^{2}_{N}$ (total), 
$B^{(IS)}(M1)$ = 0.11 $\ \mu^{2}_{N}$ and 
$B^{(IV)}(M1)$ = 12.52 $\ \mu^{2}_{N}$. 
Similarly as in the case of $^{208}$Pb, the isovector strength is larger than the isoscalar one. 
The centroid and peak energies of the full response are 
$\bar{E}^{th.}= 9.37\ \rm MeV$ and $E^{th.}_{peak} = 8.48\ \rm MeV$, respectively. 
On the other side, in the experimental investigation 
of M1 spin-flip resonance from inelastic proton scattering on 
$^{48} \rm Ca$~\cite{2016Birkhan}, 
the dominant peak locates at slightly higher energy,
 $E^{exp.}_{peak} = 10.22\ \rm MeV$. 
We note that the present IV-PV interaction, which controls the 
M1-excited state of $1^+$ configuration, is described with 
the simple, constant coupling. 
By comparing the full response with the unperturbed one,
the RRPA residual interaction shifts the main peak toward higher energy (Fig. \ref{fig:Ca01}). This is similar as shown in the $^{208}$Pb case,
demonstrating the effect of the residual RRPA interaction. 

\begin{figure}[t]
\includegraphics[scale=0.33]{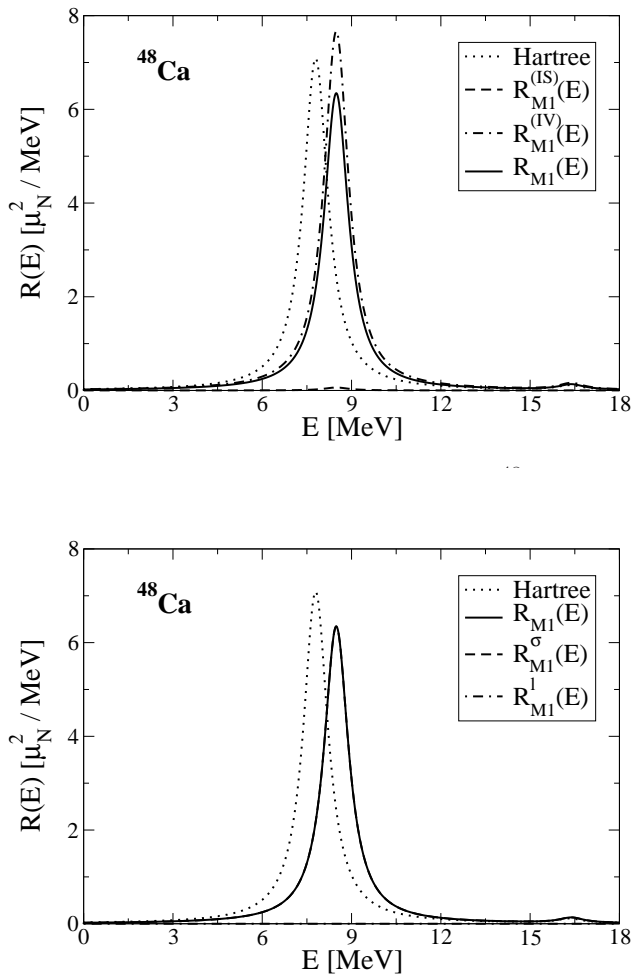}
\captionof{figure}{The same as Fig.~\ref{fig:Pb02} but for $^{48} \rm Ca$. The spin and full response functions coincide.}
\label{fig:Ca02}
\end{figure}

Figure \ref{fig:Ca02} shows the full, spin and orbital-M1 
transition strength distributions for $^{48} \rm Ca$. 
The corresponding $B(M1)$ values are $10.38~\mu^{2}_{N}$, 
$10.40~\mu^{2}_{N}$, and $5.35\times 10^{-3}~\mu^{2}_{N}$, respectively. 
Similarly as in the case of heavy system $^{208}$Pb, 
the spin transition strength dominates, i.e., it is four 
orders of magnitude larger than the orbital strength, $B^{\sigma}(M1) \gg B^{\ell}(M1)$. 

From the analysis of the major M1 state at $8.48$ MeV, we confirmed that 
it is composed mainly from the transition between the neutron 
spin-orbit partner states, $(\nu 1f^{-1}_{7/2}   \rightarrow \nu 1f_{5/2} )$. 
The major partial neutron and proton contributions 
to the B(M1) strength are presented in Table \ref{table:Th_Ca48_04}, showing 
the dominance of the neutron spin-flip transitions over the proton ones.

\renewcommand{\arraystretch}{1.5}
\begin{table}[b]
\begin{center}
\caption{Partial neutron and proton contributions $(b_{ph}^{\nu,\pi})$ to the M1 transition strength and the full transition strength $B({\rm M1})$ for the major peak at 8.48 MeV for $^{48}$Ca. 
}
\label{table:Th_Ca48_04}
\scalebox{0.95}{
\begin{tabular*}{\hsize}{ @{\extracolsep{\fill}} cccc }
 \hline  \hline
  $E^{th.}_{peak}  \lbrack \rm MeV \rbrack$   & $b_{ph}^{\nu} \lbrack \mu_{N} \rbrack$ &  $b_{ph}^{\pi} \lbrack \mu_{N} \rbrack$ &  $ B(M1) \lbrack \mu^{2}_{N} \rbrack$ \\
 \hline
      8.48  &  3.18 &  $ 4.04 \times 10^{-4}$ & 9.96\\
\hline \hline
\end{tabular*}
}
\end{center}
\end{table}
\renewcommand{\arraystretch}{1}

In the experimental investigation of M1 spin-flip resonance from 
inelastic proton scattering on $^{48} \rm Ca$~\cite{2016Birkhan}, 
the dominant peak at $E^{exp.}_{peak} = 10.22\ \rm MeV$ is 
of pure neutron character, dominated by the transition $(\nu 1f^{-1}_{7/2}   \rightarrow \nu 1f_{5/2} )$. 
This character is consistent as obtained in the present study. The measured strength is obtained as $\sum B(M1)=3.85$-$4.63~\mu^{2}_{N}$~\cite{2016Birkhan}. 
In Ref.~\cite{2016Birkhan}, the data on $^{48} {\rm Ca}(p,n)$ reaction~\cite{Yako2009} 
have also been reanalyzed, 
resulting with $\sum B(M1)=3.45$-$4.10~\mu^{2}_{N}$. 
Comparison with the RRPA results from the present analysis, $\sum B(M1)$ = 10.38 $\mu^{2}_{N}$, indicates the enhancement of the theoretical prediction for 
the $B(M1)$ transition strength. 
We note that another experimental study, based on $(\gamma, n)$ reaction~\cite{2011Tompkins}, resulted in twice as large value than in $(p,p')$ reaction~\cite{2016Birkhan}, $B(M1)= 6.8 \pm 0.5~\mu^{2}_{N}$. 
This measurement is closer, but still below the result of our present work. 
One possibility to remove this discrepancy is by introducing the quenching 
factor $\eta \simeq 0.6-0.7$ 
for the $g_{s,\ell}$ coefficients. In this case, our sum of the B(M1) strength
changes as  $\sum B(M1) \longrightarrow \eta^2 \sum B(M1) \simeq 3.7-5.1~\mu^{2}_{N}$. 
Note that the similar quenching factors have been utilized in 
several theoretical calculations \cite{Nesterenko01, Nesterenko02}. For example,
a study based on Skyrme functionals results in B(M1) values 2.5-4.8 $\mu^{2}_{N}$~\cite{Nesterenko01}.

\renewcommand{\arraystretch}{1.5}
\begin{table}[H]
\begin{center}
\caption{The peak energies $E^{th.}_{peak}$ and corresponding $B(M1)$ values for $^{42} \rm Ca$ and $^{50}$Ti in this work. 
}
\label{table:Th_Ca42_01}
\scalebox{0.95}{
\begin{tabular*}{\hsize}{ @{\extracolsep{\fill}} cccc }
 \hline  \hline
   Nuclides  &Method &$E^{th.}_{peak} \lbrack \rm MeV \rbrack$ &$B(M1) \lbrack \mu^{2}_{N} \rbrack$ \\
\hline
   $^{42}$Ca &RRPA    &  7.95    &  2.92  \\
            &RQRPA   & 11.32    & 2.12 \\
   $^{50}$Ti &RRPA    &  8.82 & 13.86 \\
            &RQRPA   &  8.58    & 8.53\\
            &        & 11.57    & 3.88 \\
\hline \hline
\end{tabular*}
}
\end{center}
\end{table}
\renewcommand{\arraystretch}{1}
%
\subsection{\label{sec:PAIRING} Pairing effects on M1 transitions}
In this section we apply the complete RHB + RQRPA framework 
adopted for the description of M1 transitions 
in open-shell nuclei, 
by consistent implementation of the pairing correlations in the nuclear 
ground state and in the RQRPA residual interaction. 
The main purpose here is to explore the role of 
the pairing correlations on the properties of the M1 response. The sensitivity of the M1 transitions on pairing correlations has previously been addressed in the study based
on the three-body model~\cite{2019OP}. In the present study, we discuss the same aspect, but utilizing a microscopic RNEDF approach. 
\begin{figure}[H]
\includegraphics[scale=0.33]{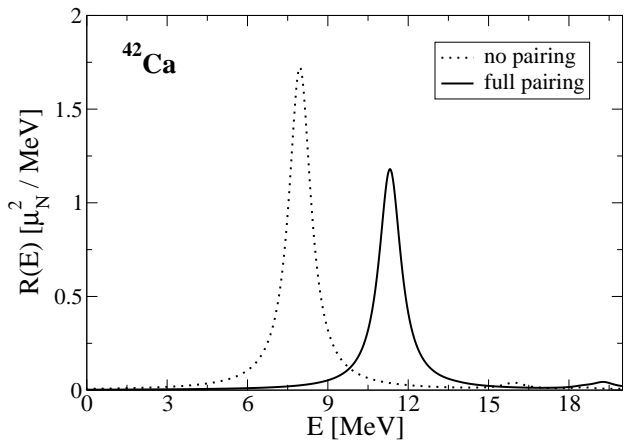}
\captionof{figure}{The RHB + RQRPA response for M1 transitions in $^{42} \rm Ca$ using the DD-PC1 parameterization and Gogny pairing correlations. The response functions without pairing correlations (RRPA) and with full pairing (RQRPA) 
are shown separately.}
\label{fig:Ca42_01}
\end{figure}
\begin{figure}[H]
\includegraphics[scale=0.33]{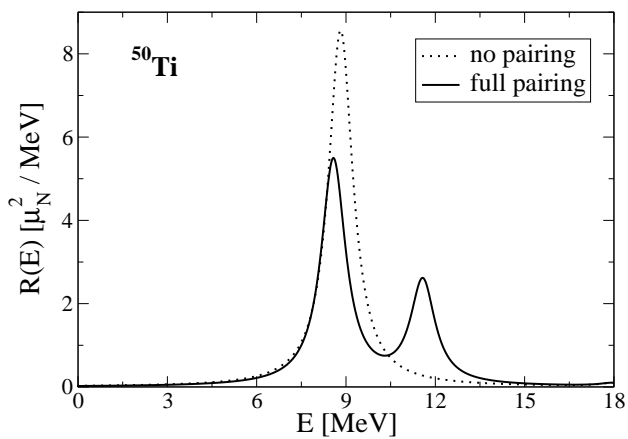}
\captionof{figure}{The same as Fig.~\ref{fig:Ca42_01} but for $^{50} \rm Ti$.}
\label{fig:Ti50_01}
\end{figure}

As the first open-shell system we study $^{42} \rm Ca$, which is already considered in its no-pairing limit in Sec.~\ref{sec:SUMRULE_TEST} in order to verify the M1 sum rule in the core-plus-two-nucleon system \cite{2019OP}. In Fig.~\ref{fig:Ca42_01}, the results of the full RHB + RQRPA calculation are shown, in comparison to the limit without pairing correlations both in the ground state (GS) and in the residual interaction. 
The full-RQRPA M1 response shows one major peak at $11.32$ MeV, 
that is composed mainly by a single-neutron $2qp$ transition, 
$(\nu 1f^{-1}_{7/2}   \rightarrow \nu 1f_{5/2} )$. 
In comparison to the peak obtained in RRPA calculation 
without pairing correlations, the pairing interaction shifts the resonant peak 
by several MeV to higher energies and reduces its strength. 
This shift by the pairing correlations at the quantitative level can also be 
seen in Table \ref{table:Th_Ca42_01}. 
The present result is consistent to that in Ref. \cite{2019OP}: 
the pairing interaction, which promotes the spin-singlet pairing between 
the valence nucleons, shifts the M1-excitation energy toward higher region, 
with a reduction of the strength, $B(M1)$. 
We note that the pairing effect in the particle-particle channel of the residual RQRPA interaction is finite but very small for the M1 transition from the $0^{+}$ GS to $1^{+}$ excited states. The major pairing effects originate from the properties of the GS. 
The same effect has been theoretically observed for the surface delta interaction applied on the M1 excitations in the particle-particle channel \cite{Tanimura01}. 

Next we move to the $^{50}$Ti nucleus, where the M1-excitations 
from its GS have been measured \cite{1985Sober}. 
These data can be utilized as reference to infer the present 
accuracy and drawback of the RQRPA calculations. 
In Fig. \ref{fig:Ti50_01}, our result is presented in the case of calculations with and without pairing correlations. The respective properties of M1 transitions are tabulated in Table \ref{table:Th_Ca42_01}. 
In the $^{50}$Ti case, the pairing interaction causes the main peak to 
split into two: $E^{th.}_{peak}=8.58$ and $11.57$ MeV. 
This two-peak structure has been confirmed in the 
experiments at $E^{exp.}_{peak} \cong 8.7$ and $10.2$ MeV \cite{1985Sober}. 
Thus, the RQRPA reasonably reproduces the general structure of M1 excitation
spectrum. 
Considering the deviation of calculated energies from the 
empirical values, further development and/or optimization of 
the RNEDF is required. 
Note also that, as we already discussed in the previous sections, 
the actual data from experiments suggests a more fine fragmentation, 
for which complicated additional effects may need to be considered, 
e.g., the meson-exchange current effect \cite{Richter01, Richter03, 1990Richter_PRL, 1994Moraghe, 2008Marcucci, Schwengner01} and/or the second-order 
(Q)RPA \cite{Dehesa, 1982Bertsch, 2006Ichimura}.

The RQRPA calculations show that the two main M1 peaks
in $^{50}$Ti mainly originate from the 
proton transition of $(\pi 1f^{-1}_{7/2}   \rightarrow \pi 1f_{5/2} )$ and 
neutron transition of $(\nu 1f^{-1}_{7/2}   \rightarrow \nu 1f_{5/2} )$.
This is because of the shell closure at $Z=N=20$, thus the main M1-excitation components are attributed to the valence, two protons and eight neutrons in the $1f_{7/2}$ orbit.

\section{\label{sec:SUM} Summary}
In this work we have introduced a novel approach to describe M1 transitions in nuclei, based on the RHB + RQRPA framework with the relativistic point-coupling interaction, supplemented with the pairing correlations described by the pairing part of the Gogny force. In addition to the standard terms of the point coupling model with the DD-PC1 parameterization, the residual R(Q)RPA interaction has been extended by the isovector-pseudovector contact type of interaction that contributes to unnatural parity transitions. 
%
A recently developed non-pairing M1 sum rule in core-plus-two-nucleon systems \cite{2019OP} has been used as a consistency check of the present theory framework. The sum of the M1-transition strength for $^{42} \rm Ca$ accurately reproduced the sum rule value (SRV), thus validating the introduced formalism 
and its numerical implementation for further exploration of M1 transitions. 

The present framework is first benchmarked on M1 transitions for two magic nuclei, $^{48}$Ca and $^{208}$Pb. The response functions $B(M1,E)$ have been explored in details, including their isoscalar and isovector components, that relate to the electromagnetic probe,  as well as contributions of the spin and orbital components of the M1 transition operator. It is confirmed that, in nuclei without deformation, the spin component of the M1 transition strength dominates over the orbital one. Due to the differences in the gyromagnetic ratios, the isovector M1 transition strength is significantly larger than the isoscalar one, and they interfere destructively. It is shown that the major peaks of isovector spin-M1 transitions are dominated mainly by a single $ph$ configuration composed of spin-orbit partner states.

One of our interests was to investigate the role of the pairing correlations on the properties of M1 response functions in open shell nuclei, that has been addressed in the study of $^{42}$Ca and $^{50}$Ti. The RQRPA calculations show a significant impact of pairing correlations on the major peak by shifting it to the higher energies, and at the same time, by reducing the transition strength. In the $^{50}$Ti case, this effect is essential to reproduce the two-peak structure measured in the experiment \cite{1985Sober}. The main effect of the pairing correlations is observed 
at the level of the ground state calculation, while it is rather small in 
the particle-particle channel of the residual RQRPA interaction.

The M1 transition strengths from the present study appear larger than the
values obtained from the experimental data. 
Therefore, it remains open question whether some additional effects should be
included at the theory side, or some strength may be missing in the experimental data. 
In addition, the M1-excitation energies of light systems, e.g. $^{48}$Ca still 
have some deviation from the reference data. In order to resolve these open questions,
further developments are needed, e.g., resolving the quenching effects in $g$ factors, meson exchange effects, couplings with complex configurations, etc.
Due to its relation to the spin-orbit interaction, M1 excitations 
could also provide a guidance toward more advanced RNEDFs.
Recently, a new relativistic energy density functional has been 
constrained not only by the ground state properties of nuclei, but also by 
using the E1 excitation properties (i.e. dipole polarizability) and giant
monopole resonance energy in $^{208}$Pb~\cite{Yuksel01}. Similarly, M1 excitation
properties in selected nuclei could also be exploited in the future studies to 
improve the RNEDFs.

\begin{acknowledgments}
This work is supported by 
the ``QuantiXLie Centre of Excellence'', a project co-financed by 
the Croatian Government and European Union through 
the European Regional Development Fund, the Competitiveness and Cohesion 
Operational Programme (KK.01.1.1.01). 
\end{acknowledgments}

\appendix
\section{Isoscalar-isovector (IS-IV) decomposition} \label{app:ISIV_DECO}
In this appendix we give some details on the IS-IV decomposition of the transition operator in a general consideration that applies both to electric and magnetic transitions.
Namely, we consider the $X\lambda \mu$ transition, where $X$ denotes $E$ or $M$ for electric and magnetic transitions, respectively, and $(\lambda,~-\lambda<\mu<\lambda)$ denote the multipole quantum numbers.
The $A$-body operator of the $X\lambda \mu$ transition is given as
\beq
\oprt{P}(X\lambda \mu)
= \sum_{i=1}^{A}  \oprt{Q}(X\lambda \mu; i),
\eeq
where $\oprt{Q}$ is the general single-particle operator.
Then, by using the isospin $\tau_3 = +1~(-1)$ for protons (neutrons),
it is expressed as
\beqa
\oprt{P}(X\lambda \mu)
&=& \sum_{k\in Z} f^{\pi}_{X\lambda} \oprt{S}(X\lambda \mu; \bir_{k})  +\sum_{l\in N} f^{\nu}_{X\lambda} \oprt{S}(X\lambda \mu; \bir_{l}), \nonumber \\
&=& \sum_{i=1}^{A}
\left[ f^{\pi}_{X\lambda} \frac{1+\hat{\tau_3}(i)}{2}
+ f^{\nu}_{X\lambda} \frac{1-\hat{\tau_3}(i)}{2} \right] \nonumber \\
&& ~~~~\cdot \oprt{S}(X\lambda \mu; \bir_{i}), \label{eq:stc25}
\eeqa
where $f^{\pi~(\nu)}_{X\lambda}$ indicates the charge (for $E\lambda$) or gyromagnetic factor (for $M\lambda$)  of the $X\lambda$ mode for protons (neutrons), and $\oprt{S}(X\lambda \mu; \bir_{i})$ is the transition operator.
See Table \ref{table:udtaf} for some examples for E2 and M1 transitions \cite{Ring01}.

\begin{table}[tb] \begin{center}
\caption{The transition operators and respective proportionality factors for E2 and M1 (orbital and spin part) modes \cite{Ring01}.
See text for details.} \label{table:udtaf}
  \catcode`? = \active \def?{\phantom{0}} 
  \begingroup \renewcommand{\arraystretch}{1.2}
  \begin{tabular*}{\hsize} { @{\extracolsep{\fill}} llll } \hline \hline
  ~$X\lambda$   &$(f^\pi,~f^\nu)$     &$(f^{IS},~f^{IV})$     &$\oprt{S}(\bir)$ \\ \hline
  ~$E2$         &$(e,~0)$           &$(e/2,~e/2)$          &$r^2Y_{2\mu}(\ubir)$\\
  &&& \\
  ~$M1,~l$      &$(g^\pi_l,~g^\nu_l)$  &$(g^{IS}_l,~g^{IV}_l)$ &$\mu_{\rm N} (\bi{\hat{l}} \cdot \bir/r)_{\mu} $ \\
  ~             &$=(1,~0)$         &$=(1/2,~1/2)$         &$\times \sqrt{3/4\pi}$ \\
  &&& \\
  ~$M1,~s$      &$(g^\pi_s,~g^\nu_s)$  &$(g^{IS}_s,~g^{IV}_s)$ &$\mu_{\rm N} (\bi{\hat{s}} \cdot \bir/r)_{\mu} $ \\
  ~             &$=(5.586,-3.826)$ &$=(0.880,~4.706)$      &$\times \sqrt{3/4\pi}$ \\  \hline \hline
  \end{tabular*}
  \endgroup
  \catcode`? = 12 
\end{center} 
\end{table}

From Eq. (\ref{eq:stc25}), the IS-IV decomposition is derived:
\beqa
&&  \oprt{P}(X\lambda \mu) = \oprt{P}^{IS}(X\lambda \mu) +\oprt{P}^{IV}(X\lambda \mu) \nonumber \\
&&= f^{IS}_{X\lambda} \sum_{i=1}^{A} \oprt{S}(X\lambda \mu; \bir_{i})  +f^{IV}_{X\lambda} \sum_{i=1}^{A} \oprt{S}(X\lambda \mu; \bir_{i}) \hat{\tau_3}(i).
\eeqa
Omitting the subscript $X\lambda$ for simplicity, the $f^{IS}$ and $f^{IV}$ factors are determined as
\beq
f^{IS}=\frac{f^\pi +f^\nu}{2},~~f^{IV}=\frac{f^\pi -f^\nu}{2},
\eeq
and equivalently, $f^\pi=f^{IS} +f^{IV}$, and $f^\nu=f^{IS} -f^{IV}$.
It is useful to check the relation between the proton-neutron
and IS-IV decompositions.
That is, by using
$\oprt{X}^\pi=\sum\limits_{k\in Z}\oprt{S}(X\lambda \mu; k)$ and
$\oprt{X}^\nu=\sum\limits_{l\in N}\oprt{S}(X\lambda \mu; l)$,
\beqa
\oprt{P}^{IS}(X\lambda \mu) &=& f^{IS} \left[ \oprt{X}^\pi + \oprt{X}^\nu \right], \nonumber  \\
\oprt{P}^{IV}(X\lambda \mu) &=& f^{IV} \left[ \oprt{X}^\pi - \oprt{X}^\nu \right],
\eeqa
where we have implemented $\tau_3=-1$ for neutrons in the last term.
Then, it is worthwhile to consider some special cases as follows.

First we assume that only the neutron component is active for the transition, namely,
$\Braket{f|\oprt{X}^\pi|i} \cong 0$ and $\Braket{f|\oprt{X}^\nu|i} \ne 0$, like as the $^{48}$Ca result in the main text.
In this case,
if $f^{IS}\ll f^{IV}$ likely as spin-M1, the
corresponding IV mode becomes dominant.
That is,
\beqa
    \abs{\Braket{f|\oprt{P}^{IS}|i}}^2 &\ll & \abs{\Braket{f|\oprt{P}^{IV}|i}}^2 \nonumber \\
 && \cong \abs{f^{IV}}^2 \abs{\Braket{f|\oprt{X}^\nu|i}}^2.
\eeqa
Otherwise,
if $f^{IS}\cong f^{IV}$ likely as E2 and orbit-M1,
\beq
 \abs{\Braket{f|\oprt{P}^{IS}|i}}^2 \cong \abs{\Braket{f|\oprt{P}^{IV}|i}}^2.
\eeq
Note also that, by considering the IS+IV response,
\beqa
  \Braket{f|\oprt{P}^{IS} +\oprt{P}^{IV}|i}
  &=& f^{\pi}\Braket{f|\oprt{X}^{\pi}|i}  +f^{\nu}\Braket{f|\oprt{X}^{\nu}|i} \nonumber  \\
  &\cong &  0+f^{\nu}\Braket{f|\oprt{X}^{\nu}|i},
\eeqa
then, this response vanishes when $f^{\nu}=0$, as in the E2 and orbit-M1 cases.
In Fig. \ref{fig:Ca02} for $^{48}$Ca, indeed the orbit-M1 response is
zero, and the total and spin-M1 results coincide.

Next, when the proton and neutron excitations occur in the same phase, it means
\beq
  \Braket{f|\oprt{X}^\nu|i} \propto (+)\Braket{f|\oprt{X}^\pi|i},
\eeq
and thus,
\beq
\abs{\Braket{f|\oprt{X}^\pi -\oprt{X}^\nu|i}}^2 \ll \abs{\Braket{f|\oprt{X}^\pi +\oprt{X}^\nu|i}}^2. \label{eq:cyoweu}
\eeq
In this case, as long as $f^{IS}\cong f^{IV}$, the IS mode is obviously dominant.
Otherwise, the result can depend on the competition of the $f^{IS}/f^{IV}$ ratio against the weightless amplitudes in Eq. (\ref{eq:cyoweu}).
Note also that, for this problem, the quantities $\Braket{f|\oprt{X}^\pi|i}$ and $\Braket{f|\oprt{X}^\nu|i}$ noticeably depend on the number of protons and neutrons, as well as the specific form of the transition operator.
In our results in the main text, for the spin-M1 mode, the IV component
is concluded as dominant commonly for the nuclides discussed.
This result is mainly attributed to that the factor $f^{IV}=g^{IV}_s$ is sufficiently larger than $f^{IS}=g^{IS}_s$.

Finally, when the proton and neutron excitations occur in the opposite phase,
$\Braket{f|\oprt{X}^\nu|i} \propto (-)\Braket{f|\oprt{X}^\pi|i}$, and thus,
\beq
 \abs{\Braket{f|\oprt{X}^\pi -\oprt{X}^\nu|i}}^2 \gg \abs{\Braket{f|\oprt{X}^\pi +\oprt{X}^\nu|i}}^2.
\eeq
In this case, the IV transition becomes dominant in the E2, orbit-M1, and spin-M1 cases, anyway.

\bibliographystyle{apsrev4-1}
%

\end{document}